\documentclass{article}[15pt]

\usepackage{amsmath,epsfig,color,calc,rotating}
\usepackage{wrapfig}

\begin{document}

\title{Partial correlation analysis indicates causal relationships between
GC-content, exon density and recombination rate in the human genome 
\vspace{0.2in}
\author{
Jan Freudengerb$^1$\footnote{Corresponding author: jan.freudenberg@nslij-genetics.org},
Mingyi Wang$^2$, Yaning Yang$^3$, Wentian Li$^1$\footnote{Corresponding author:
wli@nslij-genetics.org}\\
{\small \sl 1. The Robert S. Boas Center for Genomics and Human Genetics, 
The Feinstein Institute for Medical Research}\\
{\small \sl for Medical Research, North Shore LIJ Health System,
 Manhasset, 350 Community Drive, NY 11030, USA.}\\
{\small \sl 2. Plant Biology Division,  The Samuel Roberts Noble Foundation, Ardmore, OK 73401, USA.}\\
{\small \sl 3. Department of Statistics and Finance, University of Science and Technology
of China, Anhui 230026, Hefei, CHINA }
}
\date{preprint version of {\em BMC Bioinformatics}, 10(suppl 1):S66  (2009) }
}
\maketitle

\begin{abstract}
        % Do not use inserted blank lines (ie \\) until main body of text.
       {\bf Background:} Several features are known to correlate with 
the GC-content in the human genome, including recombination rate, gene density 
and distance to telomere. However, by testing for pairwise correlation only, 
it is impossible to distinguish direct associations from indirect ones and to 
distinguish between causes and effects. 
        {\bf Results:} We use partial correlations to construct partially 
directed graphs for the following four variables: GC-content, recombination rate, 
exon density and distance-to-telomere. Recombination rate and exon density are 
unconditionally uncorrelated, but become inversely correlated by conditioning on 
GC-content. This pattern indicates a model where recombination rate and exon 
density are two independent causes of GC-content variation.
        {\bf Conclusions:}  Causal inference and graphical models are 
useful methods to understand genome evolution and the mechanisms of isochore 
evolution in the human genome.

\end{abstract}

%%%%%%%%%%%%%%%%%%%%%%%%%%%%%%%%%%%%%%%%%%%%%%
%%                                          %%
%% The Main Body begins here                %%
%%                                          %%
%% The Section headings here are those for  %%
%% a Research article submitted to a        %%
%% BMC-Series journal.                      %%  
%%                                          %%
%% If your article is not of this type,     %%
%% then refer to the instructions for       %%
%% authors on:                              %%
%% http://www.biomedcentral.com/info/authors%%
%% and change the section headings          %%
%% accordingly.                             %% 
%%                                          %%
%% See the Results and Discussion section   %%
%% for details on how to create sub-sections%%
%%                                          %%
%% use \cite{...} to cite references        %%
%%  \cite{koon} and                         %%
%%  \cite{oreg,khar,zvai,xjon,schn,pond}    %%
%%  \nocite{smith,marg,hunn,advi,koha,mouse}%%
%%                                          %%
%%%%%%%%%%%%%%%%%%%%%%%%%%%%%%%%%%%%%%%%%%%%%%

\large
\newpage

%%%%%%%%%%%%%%%%
%% Background %%
%%
\section*{Background}

\indent

GC-content (\% of guanine(G) or cytosine(C) bases) is known to vary along human chromosomes. To describe large genomic regions of homogeneous GC\%, the term ``isochore" was coined in 1980s \cite{bernardi89}. Since then, the question has been intensively debated, why genomes contain GC-high and GC-low isochore regions. The initially proposed  hypotheses was that GC-rich isochore constitute an adaptation to homeothermy  in warm-blooded species \cite{bernardi00}, as well as favorable bendability and B-Z helix transition that lead to more open chromating and ease transcription \cite{vino03}. This explanation fits well to the correlation between GC-content and gene density \cite{ihgs01,bernardi01}. The second hypotheses to explain variation in GC-content is a mutation bias related to processes like DNA replication and repair \cite{eyre93,eyre01}. The third explanation arose from the later discovery that local GC-content and recombination rate (number of crossing over events per meiosis per unit sequence length) are strongly correlated \cite{fullerton}. The molecular basis for this explanation is recombination associated biased gene conversion (BGC), which may act to increase GC-content \cite{galtier01,eyre01,duret01,galtier03}. 

The availability of full genome sequences now allows to draw a more complex picture of GC-content variation than only separating the genome into a set of discrete isochore categories. Early after completion of the first human genome draft sequence, it was observed that seemingly homogeneous region at one length scale may not be homogeneous at shorter length scales and that it is possible to have ``domains within a domain"\cite{li01,li02}. More recently, a fine-grained picture also arose for variation of recombination rate along human chromosomes \cite{jeffreys2001,mcvean, myers2005,coop}. This facilitated the study of the relationship between GC-content and recombination rate on a much finer scale, 
showing that recombination hotspots are associated with local increases 
in GC-content, but do not significantly influence local substitution rate \cite{spencer}. In parallel, the  BGC-hypothesis has been supported by several additional lines of evidence\cite{meunier04, webster04, webster05}. In a most recent study, recombination rates was found to be the major determinant of limiting-GC-content -- the stationary GC-content towards which the human genome is currently evolving\cite{duret08}, strongly supporting recombination associated BGC as a major determinant of GC-content.

Nevertheless, it is not entirely clear how the two correlations of GC-content with both recombination rate and gene density relate to each other. In the simplest case, a third correlation between gene density and recombination rate would exist. In this case one could test whether increased GC-content in gene dense regions were a consequence of increased recombination. In the absence of a correlation between recombination rate and gene density, their shared relationship with GC-content remains to be explained. In particular, the correlation between GC-content and gene density is less understood. Thus, the true model of the evolution of genome-wide and regional GC-content may have a neutral (non-Darwinian) and additionally a (positive and negative) selection component \cite{press,bernardi07,pozzoli} or it may be void of this selection component. Because the correlation with gene density has been a major argument of evolutionary models that explain local GC-content as result of selection, a better understanding of the correlations between these variables is an important task.

To understand the relationship between recombination rate, gene density and GC-content, it is further important to note that even if BGC were the only reason for GC-content variation, this would not necessarily imply a purely neutral model of isochore evolution, because local recombination rate may itself evolve under the influence of natural selection. For instance, it has been observed that recombination is increased at human central nervous system genes and immune-system genes\cite{freudenberg07,IHC07}. These gene categories had been observed before to be subject to accelerated or faster sequence evolution, respectively\cite{ICGSC05}. Because more recombination at a genetic locus may increase the effective strength of selection, this led to the suggestion that gene selection intensity might be one determinant of local recombination rate variation \cite{freudenberg07,IHC07}.

In the present study, we aim at the assignment of ``direct" and ``indirect" labels, 
as well as ``cause" and ``effect", whenever possible, to variables that are informative about local GC-content. We notice that many previous analyses are based on statistical correlation, whereas the causal relationship between them remains undecided. For instance, researchers who are interested in understanding the causes of recombination rate variation or gene sequence evolution, GC-content itself or hidden variables associated with GC-content may be seen as possibly confounding factors. On the  other hand, for people who are interested in in GC-content variation, recombination and the associated gene conversion, and possibly mutation events, are {\sl a priori} treated as causal variables. 

When dealing with several correlated variables, a widely used statistical method is multiple regression. However, multiple regression is not always a good method to test for causal relationships, because the equality sign in a regression analysis does not have a direction. Thus, one can move an independent variable from the right-hand side of the equation to the left-hand side to be a dependent variable\cite{shipley}. Moreover, two unconditionally independent variables can be correlated conditional on a common causal child, which is exactly what is carried out in a multiple regression \cite{shipley}. Therefore, we propose to use techniques for inferring causal relationship by conditional correlation analysis to understand the relationship between GC-content, recombination rate, and gene density in the human genome. 

To this end, we start representing a group of pairwise correlated variables by an undirected graph structure: nodes/vertices represent variables and links/edges represent observed statistical correlations. In the next step, we remove all links that are inferred to be indirect associations, based on the absence of conditional correlation. Finally, we apply causal inference rules to assign causal arrows, if possible. In cases where the complete causal model cannot be inferred from the data, the result is a partially directed graph that optimally characterizes the relationship among the tested variables. Similar inference techniques have been previously applied to other genomics problems\cite{fuente} and for studying relationships between human-disease related intermediate-phenotypes\cite{li06}.

%%%%%%%%%%%%%%%%%%%%%%%%%%%%
%% Results and Discussion %%
%%
\section*{Results and discussion}

\subsection*{Three variables: GC\%, recombination rate, and 
distance to telomere}

\indent

In a recent study, it was shown by Arndt and Duret\cite{duret08} that besides the positive correlation with recombination rate (RR), GC-content (GC\%) is negatively correlated with the distance to telomere (DT). These results were mainly based on the analysis of noncoding sequence in a 1Mb sized window that have high quality finished sequence available both in the chimpanzee and the macaque genome\cite{duret08}. We start our analysis by using both their data and our own dataset of the same 1Mb windows for the human genome sequence, regardless of coding and noncoding status or the existence of quality sequence in other organisms. The GC\% in these two datasets is not totally identical, but highly correlated ($\rho=0.98$). Similarly, the HapMap estimate of RR \cite{IHC07} in the two datasets is correlated with $\rho=0.82$. We discarded windows, if the number of HapMap single-nucleotide-polymorphism (SNP) is less than 20 or more than 30\% of genomic sequence are missing. In total, 2647 and 2668 1Mb windows are available with information on GC\%, RR and DT for the two datasets. We performed log-transformation of distance to telomere (DT), because the scatter plot showed a non-linear correlation between DT with the other two variables, and then multiplied it by $-1$ to change the negative correlation with GC\% to positive. The unconditional and conditional Pearson's correlation coefficients between GC\%, RR and DT are shown in Table 1. All correlation coefficients are highly significant ($p$-value=0) and results from both datasets are highly similar.

Because an earlier study had observed that the correlation between RR and GC\% is maximal when both variables are measured in the 50kb window\cite{mcvean}, we also looked at a dataset where GC\%, RR, DT are measured by using the window size of 50kb. Due to the smaller window size (1/20 of  the 1Mb window), RR is fluctuating in a much wider range as can be seen from the quantile values in Table 2. We also note that a square-root transformation of RR under 50kb window leads to a slightly better linear correlation with GC\%, and a larger correlation coefficient (result not shown).  

The correlation and partial correlation between the three variables from 50kb window is shown in Table 3. In contrast to \cite{mcvean}, we found the correlation between GC\% and RR to be higher using the 1Mb sized window than the 50kb window. This discrepancy may result from the threefold higher SNP density provided by the HapMap phase II \cite{IHC07}. Importantly, the correlation between GC\% and DT is less affected by the change of window size, although RR-DT correlation is far weaker in the 50kb window than in the 1Mb window. This change of the strength of the correlation of RR with GC\% and DT from one window size to another may be related to the ``domains within  domains" phenomenon that had been found for GC-content variation and that may exist for fine-scale recombination rate variation too. 

Because none of the pairwise correlations between GC\%, RR and DT is rendered insignificant by conditioning on the third variable, it is not possible to remove any edge in the relationship graph for GC\%, RR, and DT (Figure 1(A)).

\subsection*{Chromosome-specific correlation and partial correlations}

\indent

In the next step, we checked the chromosome-specific correlations and partial correlations between the three variables. Table 4 shows these result in form of correlation and partial correlation coefficients (and $p$-value if it is larger than 0.01) for our main dataset (1Mb window including all available human genome sequence independent from its coding status). There are several notable observations:  (1) RR-$\log(1/DT)$ correlation is unchanged by conditioning on GC\%  for non-acrocentric chromosomes, indicating that the position of the window already explains RR, rendering GC\% unlikely to be causal. (2) For acrocentric chromosomes (13, 14, 15, 21, 22), the position of the window (DT) is only marginally correlated with RR. In contrast, DT is correlated with GC\% for all chromosomes including the acrocentric chromosomes. (3) For some (3, 7, 8, 9, 10, 11, 12, 18, 19), but not for all, chromosomes, the correlation between GC\% and RR is weakened by conditioning on DT.(4) For chromosome 2 the positive correlation between RR and DT is not turned negative by conditioning on GC. This result is interesting, because chromosome 2 is known to result from a relatively recent fusion event of different chromosomes \cite{dreszer07, ijdo91}

To examine the robustness of these chromosome specific correlations (Table 4), we carried out the same correlation analysis using the noncoding sequence 1Mb windows\cite{duret08} and the 50kb  window (Figure 2). Most of the correlations in Table 4 are confirmed in these two additional datasets. One interesting observation in Figure 2 is that the correlation between RR and DT is  weaker for the 50kb window, probably because finer details of recombination rate variation are revealed at this length scale and the dependence of RR on DT is no longer monotonic. Thus DT is primarily correlated with large scale recombination rate variation, which could relate to the proposed conservation of large-scale rates on longer time scales \cite{duret08,mcvean}.

An example of chromosome specific patterns of recombination rate was recently discussed in the context of a putative gene that controls overall recombination rate\cite{kong08}. This study illustrates the effect of a SNP on increasing the female recombination rates by almost the same amount on all chromosomes with the exception of chromosome 21. Another SNP reduces the male recombination rates by variable degrees for different chromosomes\cite{kong08}.

\subsection*{Three variables: GC\%, recombination rate, and number of exons}

\indent

Gene density constitutes a further variable that is known to be strongly correlated with GC\% \cite{ihgs01,bernardi01}. To better understand this relationship, we counted the number of exons within a 1Mb window, as it reflects both the number of genes and the intron count. The correlation and partial correlations between GC\%, RR, and the number of exons (NE) are listed in Table 5. Unlike the previous situation, where we had looked at the three variables RR, DT and GC\%, the consideration of NE instead DT is bringing up an observation that allows us to infer a causal relationship: although no significant direct correlation exists between RR-NE, a negative correlation between RR and NE emerges after conditioning on GC\%. 

This result (Table 5) suggests the causal model in Figure 1(B). In this causal model, RR and NE are two independent causes of GC\%. The inference of this causal structure is based on the known fact that conditioning on a common child variable creates a correlation between two previously uncorrelated causes of this child variable \cite{shipley}. Or spoken more specifically, the relationship between NE and RR can be understood as follows: normally the two variables RR and NE do not contain any information about each other and are therefore uncorrelated. However, given the status of GC-content as third variable, this situation changes and RR and NE are now mutually informative. This mutual informativeness of NE and RR depending on GC\% is explained by a model where both RR and NE are independent causes of GC\%. When GC\% in a region is high and RR is low, NE is more likely to be high. Vice versa, when NE is low, RR is more likely to be high. Thus, given the status of GC\%, a previously invisible relationship between RR and NE emerges due to the causal influence of both variables on GC\%.

Consistent with our present observation, a negative correlation between gene density and RR had been observed earlier in a multiple regression analysis when looking at 3Mb windows, despite the fact that the unconditional RR/gene count correlation was weakly positive \cite{kong}. Importantly, window size could be a factor that exerts some influence on the magnitude of observed correlations. Recombinations tend to occur more often in physical proximity to genes, when compared to intergenic regions; but on the other hand, they also tend to occur away from exons on a finer scale\cite{coop}. It might be due to this subtle variation of RR at different length scales that the correlation between RR-NE is insignificant at the 1Mb scale, but was weakly positive on the 3Mb scale.

Nevertheless, when we repeated the chromosome-specific analysis using the variable NE (instead of DT), this confirmed the overall pattern of correlation between RR and NE. Unconditionally the correlation is not significant and can be both positive and negative. However, the partial correlations between NE and RR conditional on GC\% are all negative with most of them being significant (results not shown).

In principle, the absence of an unconditional correlation between RR and NE could also result from a 
phenomenon termed suppression \cite{cramer,lewis,tu}. Suppression refers to the situation, where different signs are obtained by following two paths with opposite effects from the same starting to the same ending node. However, the observed change of the correlation from insignificant to significant is inconsistent with suppression, because this conditional dependence indicates that both the NE and the RR link with GC\% are pointing towards GC\%.

\subsection*{Four variables: GC\%, recombination rate, distance to telomere,
and number of exons}

\indent

In the final step, we extended our 3-variable analysis to a 4-variable analysis, which includes GC\%, RR, $-\log(DT)$, and NE. Besides the previously calculated first-order partial correlation (conditional on one variable), we now also calculate the second-order partial correlations (conditional on two other variables). The result are shown in Table 6. When comparing the second-order partial correlations to the first-order partial correlations, we found that conditioning on GC\% is mostly responsible for any change of correlation status. Conditioning on DT, RR or NE has only some quantitative effect, instead of introducing any qualitative changes into pairwise and first-order correlations. This implies a central position of GC\% among these variables.
\indent

Figure 1(C) depicts a partially directed graph that is consistent with the results in Table 6. Importantly, the inclusion of DT does not alter causal relationships that were inferred above in the 3-variable analysis of RR, NE and GC\%. Also, the above correlations between RR, DT and GC\% remain largely unaltered by the inclusion of NE. As mentioned above, telomere distance is inversely correlated with GC\% and RR. Additionally, we see in the unconditional pairwise analysis that telomere distance is inversely correlated with NE too, although this correlation is of smaller magnitude. This correlation between DT and NE does not change substantially when conditioning on RR. However, when conditioning on GC\%, the correlation between DT and NE changes its direction. Following a similar line of reasoning as above, this suggests a model where DT and NE are two independent causes of GC\%. On the contrary, this cannot be said for the influence of RR and DT on GC\%, because the correlation RR and DT does not depend on conditioning on GC\%. 

To find the missing orientations of the links between RR, DT and GC\% in the 4-variable model, we next applied the TETRAD program\cite{tetrad} that implements the PC-algorithm to create a causal model by a systematic search strategy (see Methods for details) \cite{pc, spirtes}. The graphical result that we obtained from running TETRAD is essentially the same as the one depicted in Figure 1(C) and confirmed the direction of the two arrows that we had inferred for causative influence of both RR and NE on GC\%. However, the additionally proposed orientations of the links RR $\rightarrow -\log(DT)$ and NE $\rightarrow -\log(DT)$ are biologically counterintuitive, because telomere distance is unlikely to be an effect of any of the other variables. 

To explain the difficulty to infer the directions of these causal links between RR, DT and GC\%, we hypothesize the causal model in Figure 1(D). This model includes as fifth hidden variable the proportion of recombination events that are resolved exclusively as gene conversion event without any crossing-over event (NCO/R), a variable that was recently suggested to be important \cite{duret08}. In this model in Figure 1(D), NCO/R is a cause of GC\% that does not fully depend on RR, but is influenced in its magnitude by RR. A similar relationship might connect NCO/R with NE. On the other hand, distance-to-telomere, similar to other variables measuring position or time, might play the role of providing a common environment for several other variables. In other words, one can draw a directed arrow from DT to all other variables under discussion. A similar situation is seen for the linkage disequilibrium between two neighboring genetic markers, where the position can be considered is a ``cause" of both markers. However, we could not test the validity of the model in Figure 1(D) because NCO/R data are not available.

%%%%%%%%%%%%%%%%%%%%%%
\section*{Conclusions}

\indent

We apply partial correlation and graphical probabilistic model inference to several genomic variables that are correlated with GC-content in the human genome. We can show that recombination rate and exon density are two independent causes of GC\% as measured on the 1Mb scale. This observation adds some support to models that complement the influence of recombination rate on GC-content with a component involving selection. In addition, it appears unlikely that GC\% variation is a cause of variation in recombination rate or exon density. We observe some heterogeneity in the human genome, such as differences in the correlation of RR with the distance to telomere between acrocentric and non-acrocentric chromosomes. We also see indications of window-size dependent correlation pattern, which may reflect the subtle differences of the distribution of recombination near and within genes.

%%%%%%%%%%%%%%%%%%
\section*{Methods}

\subsection*{Terminology in relationship and causal graphs}

\indent

A graph G=(V,E) contains vertex/node set V and edge/link  set E $\subseteq V \times V$.  An edge ($i,j$) $\in E$ is  ``directed" if ($j, i$) $\notin E$; and is ``undirected" if  ($j, i$) $\in E$. If there is an edge between node $i$ and $j$,  either directed or undirected, we say there is a ``direct  association/relationship" between the two nodes. If there is no edge between node $i$ and node $j$, the two are still connected through multiple-step edges, as all our nodes are in one single graph; then we say the two nodes are ``indirectly associated". 

If all edges are directed, the graph is said to be ``directed graph" (e.g. Figure 1(B)). If all edges are undirected, the graph is an ``undirected graph" (e.g. Figure 1(A)). If some edges are directed and other edges are undirected, the graph is a ``partially directed graph" (e.g. Figure 1(C)).

\subsection*{Partial correlations}

\indent

For many situations, conditional correlation is equivalent to partial correlation\cite{baba} which is defined as follows (with one control variable $z$):
\begin{equation}
\label{eq1}
\rho_{xy.z} = \frac{ \rho_{xy} - \rho_{xz} \rho_{yz}}{ \sqrt{(1-\rho_{xz}^2)(1-\rho_{yz}^2)}}.
\end{equation}
where $z$ ($\rho_{xy}=cov(x,y)/\sqrt{ var(x) var(y)}$ is the Pearson product-moment correlation coefficient. From the linear regression framework, partial correlation is the
correlation after the main terms in regression over $z$ are removed:
\begin{eqnarray}
\label{eq2}
x &=& a_x+b_x z + \epsilon_x  \nonumber\\
y &=& a_y+b_y z + \epsilon_y  \nonumber \\
\rho_{xy.z}& = & Cor( \epsilon_x, \epsilon_y)
\end{eqnarray}
Partial correlation $\rho_{xy.z}$ is often lower than $\rho_{xy}$, and a significantly lower partial correlation is an indication that the $x-y$ correlation is indirect.

With more than 3 variables ($x,y,z,w$), the partial correlation can be defined by conditional on one variable (e.g. $z$, first order), or two variables ($z,w$, second order). Both Eq.(\ref{eq1}) and Eq.(\ref{eq2}) can be extended for calculating second-order partial correlation:
\begin{equation}
\label{eq3}
\rho_{xy.zw} = \frac{ \rho_{xy.z} - \rho_{xw.z} \rho_{yw.z}}{\sqrt{(1-\rho_{xw.z}^2)(1-\rho_{y
w.z}^2)}}
\end{equation}
and
\begin{eqnarray}
\label{eq4}
x &=& a_x+b_x z +c_x  w + \epsilon_x  \nonumber\\
y &=& a_y+b_y z + c_y w +  \epsilon_y  \nonumber \\
\rho_{xy.zw}& = & Cor( \epsilon_x, \epsilon_y).
\end{eqnarray}

Higher order partial correlation can be defined in an analog fashion.

\subsection*{Establishing undirected, partially directed and directed graphs}

\indent

Figure 3 illustrates an example for inferring relationship and causal graph from data for three variables $x,y,z$. In Figure 3(A), we assume all pairwise correlations are significant, so all nodes are linked to other nodes. If the correlation between two of the variables is not due to a direct cause-effect relationship, but mediated via a third variable, then the correlation between the two conditional on that third variable will be greatly reduced. Accordingly, we would end up Figure 3(B), if assuming that the partial correlation $Cor(x,y|z)$ becomes insignificant, while the other two partial correlations remain significant. In that case, partial correlation cannot determine the orientation of causal arrows. Except the first causal model on the top of Figure 3(C), all the three other causal models were possible. However, in a special situation, a directed causal model can be inferred uniquely. Suppose $Cor(x,z)$ and 
$Cor(y,z)$ are both significant, but $Cor(x,y)$ is insignificant, then we start with the undirected graph in Figure 3(B) from the unconditional analysis. Further suppose $Cor(x,y|z)$, 
$Cor(x,z|y)$, $Cor(y,y|x)$ are all significant. Then by the rule of d-separation \cite{spirtes} only the top model in Figure 3(C) is consistent with these assumptions.

\subsection*{TETRAD program for inference of causal models from partial correlations}

\indent

The TETRAD program\cite{tetrad} implements the PC-algorithm to automatically infer 
causal relationships from partial correlation analysis \cite{pc}. This algorithm 
can be broken into two phases: an adjacency phase and an orientation phase. 
In the adjacency phase, a complete undirected graph over the variables is constructed 
and then edges $X-Y$ are removed, if some set $S$ among either the adjacents of $X$ or 
the adjacents of $Y$ can be found such that $I(X, Y | S)$. Then the orientation 
phase is begun. The first step examines unshielded triples and considers to orient 
them as colliders. An unshielded triple is a triple $(X, Y, Z)$ where $X$
is adjacent to $Y$, $Y$ is adjacent to $Z$, but $X$ is not adjacent to $Z$. 
Since $X$ is not adjacent to $Z$, the edge $X-Z$ must have been removed during the 
adjacency search  by conditioning on some set $S_{xz}$; $(X, Y, Z)$ is oriented as 
a collider $X \rightarrow Y \leftarrow Z$ just in case $Y$ is not in this 
$S_{xz}$. Once all such unshielded triples have been oriented as colliders, a 
series of rules orients any edge whose orientation is implied by previous orientations.

%%%%%%%%%%%%%%%%%%%%%%%%%%%%%%%%
\section*{Authors contributions}

W.L. proposed the project, carried out the correlation and partial correlation calculation, 
wrote the initial draft of the manuscript; J.F. prepared the data, contributed most 
of the biological discussion, wrote the final version of the manuscript; 
M.W. ran the TETRAD program, contributed to the theoretical aspect of causal inference; 
Y.Y. contributed to the theoretical aspect of partial correlation.

%%%%%%%%%%%%%%%%%%%%%%%%%%%%%%%%
\section*{List of abbreviations}

{\bf BGC:} biased gene conversion,
{\bf DT:} distance to telomere,
{\bf GC\%:} guanine and cytosine content,
{\bf NCO/R:} non-crossing-over events among recombination events,
{\bf NE:} number of exons,
{\bf PC-algorithm:} Peter (Spirtes) and Clark (Glymour) algorithm,
{\bf RR: } recombination rate,
{\bf SNP:} single nucleotide polymorphism.

%%%%%%%%%%%%%%%%%%%%%%%%%%%
\section*{Acknowledgements}

We thank Peter Arndt for sharing the data he used in his paper.
Yaning Yang was supported by Chinese Natural Science Foundation (No. 10671189).

\newpage

%%
%% Do not use \listoffigures as most will included as separate files

\section*{Figures}
  \subsection*{Figure 1 - Causal graph models or their skeleton 
for GC-content, recombination-rate, number-of-exons, and distance-to-telomere}
(A) Relationship graph for GC\%, RR, $-\log(DT)$ that
is inferred from the correlations in Table 1. (B) Causal graph for
GC\%, RR, NE that is inferred from the correlations in Table 5. 
(C) Partially directed graph for GC\%, RR, $-\log(DT)$, and NE
that is consistent with the result in Table 6. All edges/arrows are 
highly significant. (D) A hypothetical model including an extra 
variable NCO/R: proportion of non-crossing-over events. This model 
may help to orient the previously undirected edges.

\begin{figure}[t]
 \begin{center}
  \begin{turn}{-90}
   \epsfig{file=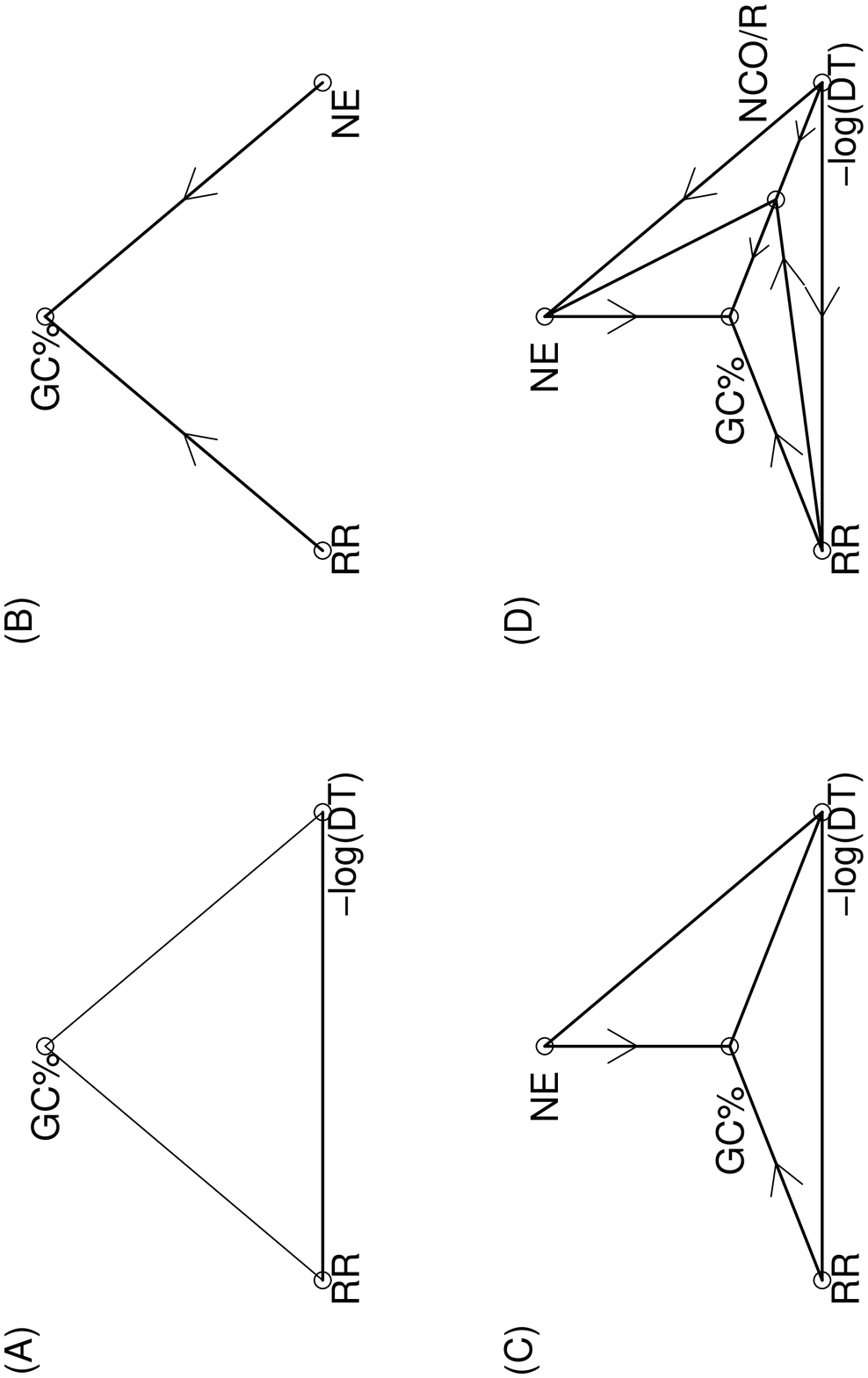, width=10cm}
  \end{turn}
 \end{center}
\end{figure}

\newpage
  \subsection*{Figure 2 - Chromosome-specific correlations and
partial correlations}
Chromosome-specific correlation and
partial correlation for GC\%-RR (top), GC\%-$\log(1/DT)$ (middle),
 RR-$\log(1/DT)$ (bottom) in 3 datasets: 1Mb, non-coding
(black)\cite{duret08}; 1Mb, disregard coding/non-coding status (blue);
and 50kb, disregard coding/non-coding status (red).
Acrocentric chromosomes are marked by yellow bars.

\begin{figure}[t]
 \begin{center}
  \begin{turn}{-90}
   \epsfig{file=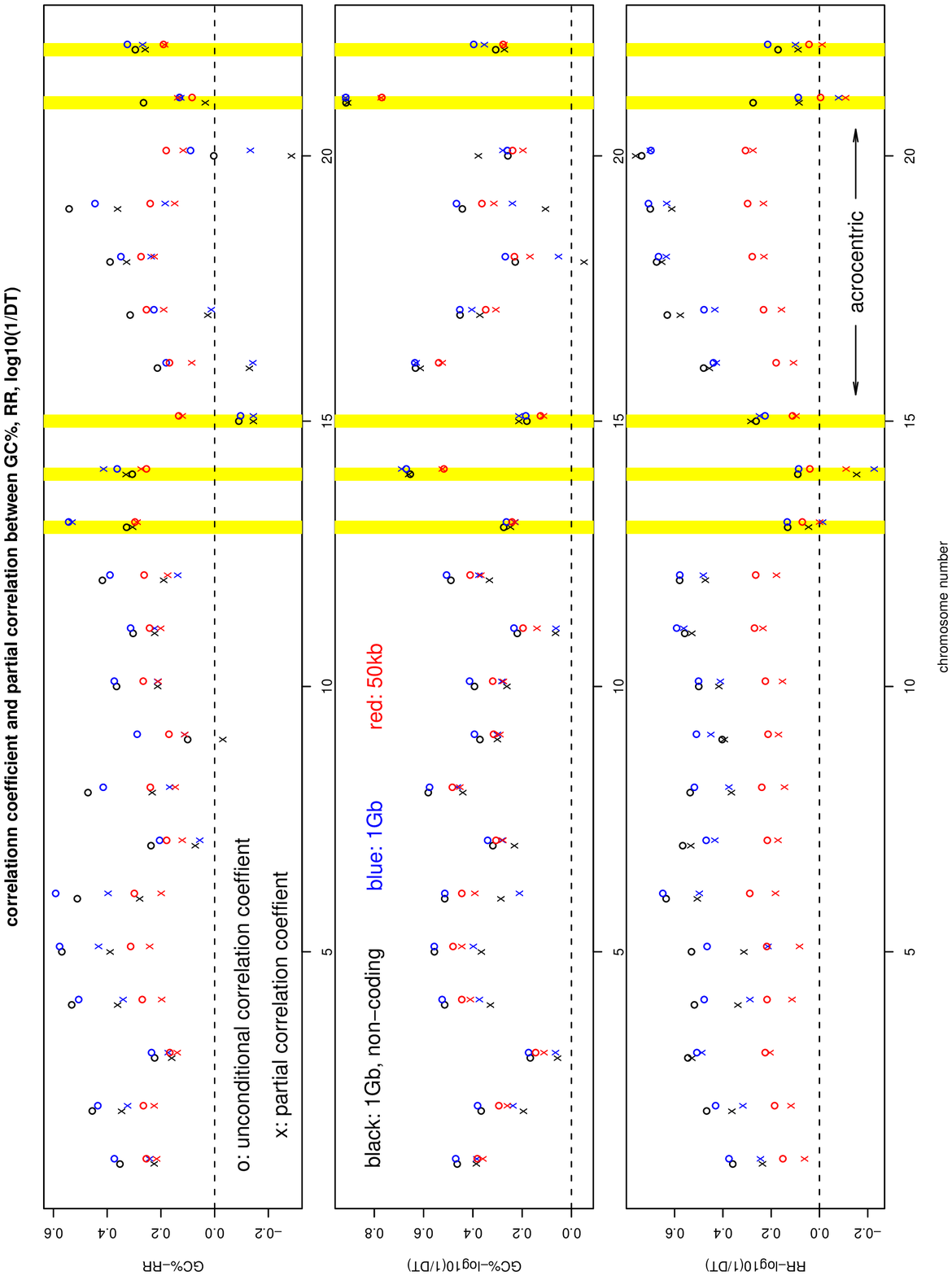, width=12cm}
  \end{turn}
 \end{center}
\end{figure}

\newpage
  \subsection*{Figure 3 - Illustration of the procedures in
establishing undirected, directed, or partially directed graphs. }
(A) If correlation $Cor(i,j)$ is significant, draw
an edge between node $i$ and node $j$ ($i,j \in (x,y,z)$). 
(B) The conditional correlation $Cor(x,y|z)$ is insignificant,
remove the edge $(x,y)$.
(C) Using other information to select one or few causal
models that are consistent with the data.

\begin{figure}[t]
 \begin{center}
  \begin{turn}{-90}
   \epsfig{file=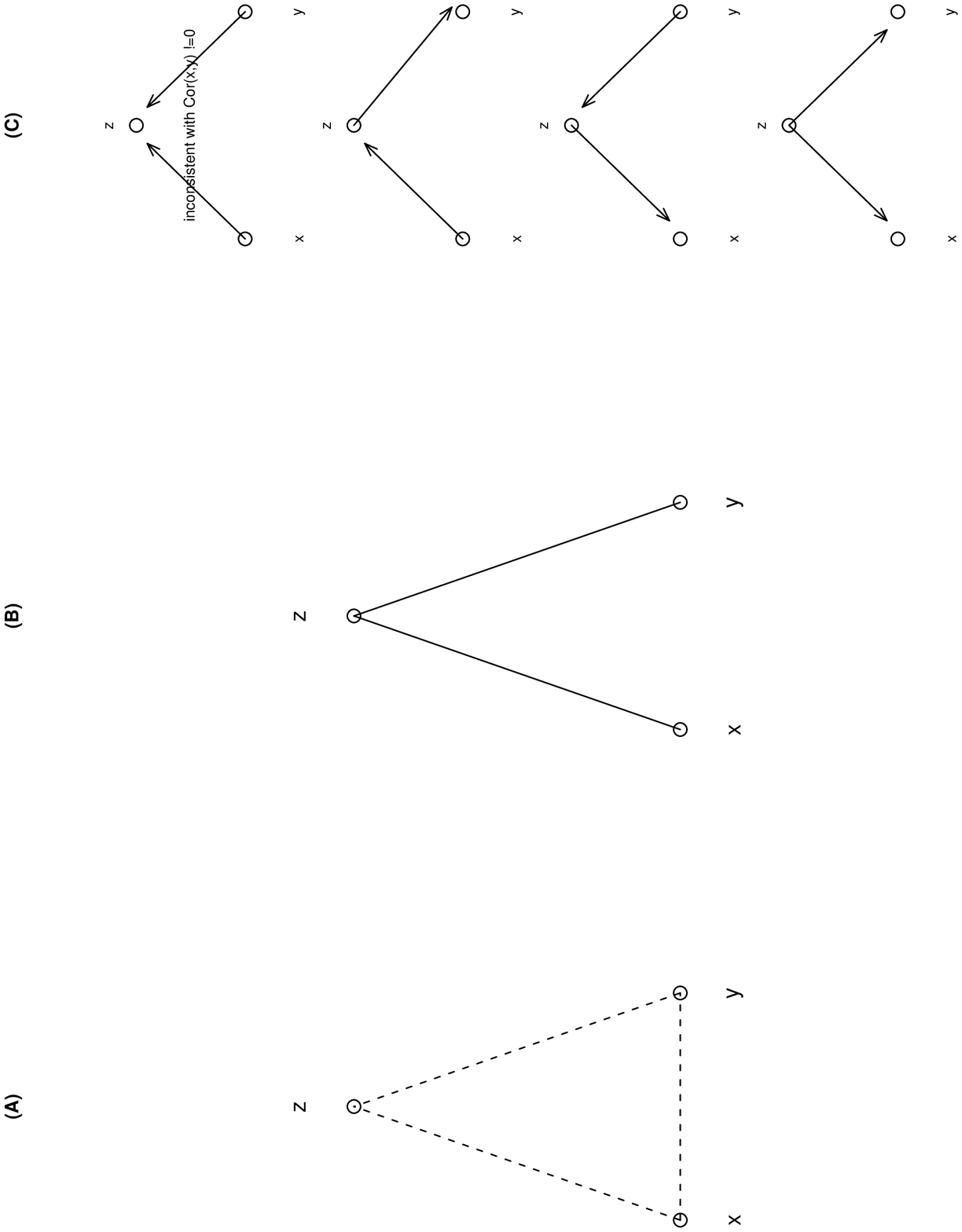, width=11cm}
  \end{turn}
 \end{center}
\end{figure}

\newpage

%%%%%%%%%%%%%%%%%%%%%%%%%%%%%%%%%%%
%%                               %%
%% Tables                        %%
%%                               %%
%%%%%%%%%%%%%%%%%%%%%%%%%%%%%%%%%%%

%% Use of \listoftables is discouraged.
%%
\section*{Tables}

\subsection*{Table 1 - Correlation and and partial correlation at 1Mb windows. }

Correlation and and partial correlation between GC\%, recombination rate (RR), 
and distance to telomere (DT) (negative log-transformed) for 1Mb windows. 
Conditioning is performed on the respective third variable.
(A) regardless of coding status; (B) non-coding only. \\

 (A)\\
 \begin{tabular}{c|cc}
 \hline
 $\rho$/partial-$\rho$ &  RR & $-\log(DT)$ \\
 \hline
 GC\% &  0.38/0.20 & 0.47/0.35 \\
 RR & &  0.49/0.38 \\ 
 \hline
 \end{tabular}

 (B)\\
 \begin{tabular}{c|cc}
 \hline
 $\rho$/partial-$\rho$  & RR & $-\log(DT)$ \\
 \hline
 GC\% &  0.39/0.20 & 0.46/0.33 \\
 RR &  &  0.52/0.42\\
 \hline
 \end{tabular}

\subsection*{Table 2 - Quantile values of recombination rates}
Quantile values for RR for the three datasets: 1Mb non-coding, 
1Mb and 50kb (in cM/Mb). \\

\begin{tabular}{c|ccccc}
\hline
dataset & 0\% & 25\% & 50\%  & 75\% & 100\%\\
\hline
1Mb,nc   & .012 & .80 & 1.19 & 1.82 & 4.97 \\
1Mb      & .033 & .90 & 1.40 & 2.10 & 7.47 \\
50kb     & 0    & .26 & 0.72 & 1.89 & 27.55 \\
\hline
\end{tabular} 

\newpage

\subsection*{Table 3 -
Correlation and partial correlation at 50kb windows}
Correlation and partial correlation between GC\%, RR, and DT 
(negative log transformed) for 50kb windows.\\

\par 
\mbox{

\begin{tabular}{c|cc}
\hline
 $\rho$/partial-$\rho$ & RR & $-\log(DT)$ \\
\hline
GC\% &  0.25/0.17 & 0.40/0.36  \\
RR  &             & 0.22/0.14 \\
\hline
\end{tabular}
}

\subsection*{Table 4 - Chromosome-specific correlation and partial correlation}

Chromosome-specific correlation and partial correlation between GC\%, 
RR, and DT (negative log-transformed) using the 1Mb window.  
A $p$-value for testing zero-correlation is included only when the correlation 
is not significant.  $n$ is the number of windows per chromosome 
(i.e., sample size). Acrocentric chromosomes are marked by *.

\par
\mbox{

\begin{tabular}{c|ccc|c} \hline chr & GC\%-RR & GC\%-$\log(1/DT)$ & RR-$\log(1/DT)$ & $n$\\ & \multicolumn{3}{c|}{ $\rho$ ($p$-value)/ partial-$\rho$ ($p$-value) } & \\ \hline
1 & 0.37/0.24 & 0.47/0.38 &0.37/0.24 & 224 \\
2 & 0.43/0.32 & 0.38/0.24 &0.43/0.32 & 238 \\
3 & 0.23/0.17(0.016) & 0.17(0.015)/0.065(0.37) &0.51/0.49 & 194 \\
4 & 0.51/0.34 & 0.52/0.37 &0.48/0.29 & 187 \\
5 & 0.58/0.43 & 0.56/0.40 &0.47/0.21 & 175 \\
6 & 0.29/0.40 & 0.51/0.21 &0.65/0.50 & 166\\
7 & 0.20(0.01)/0.054(0.50) & 0.34/0.28 &0.47/0.43 & 154\\
8 & 0.41/0.17(0.048) & 0.58/0.46 &0.52/0.38 & 142\\
9 & 0.29/0.11(0.24) & 0.39/0.30 &0.51/0.45 & 114 \\
10 &0.37/0.21(0.015) & 0.41/0.28 &0.50/0.41 & 131 \\
11 &0.31/0.22(0.01) & 0.23/0.063(0.48) &0.59/0.56 & 130\\
12 &0.39/0.14(0.12) & 0.51/0.37 &0.58/0.48  & 129\\
13* &0.54/0.53 & 0.26/0.23(0.025) &0.13(0.20)/$-$0.012(0.91) & 95 \\
14* &0.36/0.41 & 0.67/0.69 &0.086(0.42)/$-$0.23(0.035) & 87 \\
15* &-0.097(0.38)/-0.14(0.19) & 0.19(0.096)/0.21(0.054) &0.22(0.04)/0.25(0.024) & 82\\
16 &0.18(0.11)/-0.14(0.22) & 0.64/0.63 &0.44/0.43 & 77\\
17 &0.22(0.04)/0.012(0.92) & 0.45/0.40 &0.48/0.43 & 77\\
18 &0.35/0.24(0.04) & 0.27(0.02)/0.051(0.66) &0.67/0.63 & 74\\
19 &0.45/0.18(0.18) & 0.47/0.24(0.082) &0.71/0.63  & 54 \\
20 &0.089(0.50)/-0.13(0.31) & 0.26(0.046)/0.28(0.033) &0.70/0.70 &59\\
21* &0.13(0.48)/0.13(0.50) & 0.92/0.92 &0.088(0.61)/$-$0.079(0.67) &32\\
22* &0.32(0.06)/0.27(0.13) & 0.40(0.02)/0.35(0.04) &0.21(0.22)/0.099(0.58) &34 \\
\hline
ave& 0.38/0.20 & 0.47/0.35 & 0.49/0.38 & 2655 \\
\hline
\end{tabular} 

}

\newpage

\subsection*{Table 5 - adding number-of-exon variable}

Correlation and partial correlation between GC\%, RR, and number of exons 
(NE) in 1Mb windows.\\

\par
\mbox{
 \begin{tabular}{c|ccc}
 \hline
 $\rho$/partial-$\rho$  & RR & NE \\
 \hline
 GC\% & 0.38/0.49 & 0.69/0.73  \\
 RR   &           & 0.04/-0.34  \\ 
 \hline
 \end{tabular}
}

\subsection*{Table 6 - Correlation and and partial correlation between four variables: GC\%, RR, DT and NE}

In addition, the first-order partial correlations for RR-NE and DT-NE pairs are 
shown, whereas the first order partial correlations between the other variables 
had been already shown above.

\par
\mbox{
 \begin{tabular}{c|ccc}
 $\rho$/partial-$\rho$ & RR & $-\log(DT)$ & NE \\
\hline
 GC\% & 0.38/0.33 & 0.47/0.33 & 0.69/0.72 \\
\hline
 RR   &           & 0.49/0.33 & 0.036/-0.28 (cond. on GC\% and log(1/DT)) \\
 &           &             & /-0.34 (cond. on GC\%) \\
 &           &             & /-0.056 (cond. on log(1/DT) ) \\
\hline
 $\log(1/DT)$ & &             & 0.17/-0.12 (cond. on GC\% and RR) \\
              & &             & /-0.23 (cond. on GC\%)  \\
              & &             & /0.18 (cond. on RR)  \\
 \hline
 \end{tabular}
}

\end{document}